\documentclass[10pt,twocolumn,letterpaper]{article}

\usepackage[breaklinks=true, bookmarks=false]{hyperref}
\usepackage{cvpr}
\usepackage{times}
\usepackage{epsfig}
\usepackage{graphicx}
\usepackage{dsfont}
\usepackage{amsmath}
\usepackage{amssymb}
\usepackage{comment}
\usepackage{pifont}
\usepackage{multirow}
\DeclareMathOperator{\E}{\mathbb{E}}
\newcommand{\cmark}{\ding{51}}
\newcommand{\xmark}{\ding{55}}

\cvprfinalcopy 

\def\cvprPaperID{24} 

\ifcvprfinal\fi
\begin{document}

\title{Structure Preserving Compressive Sensing MRI\\ Reconstruction using Generative Adversarial Networks}

\author{Puneesh Deora$^1$\thanks{Equal contribution} \quad Bhavya Vasudeva$^1$\footnotemark[1] \quad Saumik Bhattacharya$^2$ \quad Pyari Mohan Pradhan$^1$\\
$^1$Dept. of ECE, IIT Roorkee, Uttarakhand, India\\
$^2$Dept. of E\&ECE, IIT Kharagpur, West Bengal, India\\
{\tt\small \{pdeora,bvasudeva\}@ec.iitr.ac.in \quad saumik@ece.iitkgp.ac.in\quad pmpradhan.fec@iitr.ac.in}
}

\maketitle
\thispagestyle{empty}

\begin{abstract}
Compressive sensing magnetic resonance imaging (CS-MRI) accelerates the acquisition of MR images by breaking the Nyquist sampling limit. In this work, a novel generative adversarial network (GAN) based framework for CS-MRI reconstruction is proposed. Leveraging a combination of patch-based discriminator and structural similarity index based loss, our model focuses on preserving high frequency content as well as fine textural details in the reconstructed image. Dense and residual connections have been incorporated in a U-net based generator architecture to allow easier transfer of information as well as variable network length. We show that our algorithm outperforms state-of-the-art methods in terms of quality of reconstruction and robustness to noise. Also, the reconstruction time, which is of the order of milliseconds, makes it highly suitable for real-time clinical use.
\end{abstract}
%
%
\section{Introduction}
\label{sec:intro}

Magnetic resonance imaging (MRI) is a commonly used non-invasive medical imaging modality that provides soft tissue contrast of excellent quality as well as high resolution structural information. The most significant drawback of MRI is its long acquisition time as the raw data is acquired sequentially in the k-space which contains the spatial-frequency information. This slow imaging speed can cause patient discomfort, as well as introduce artefacts due to patient movement.    
\begin{figure}[h!]
\center
    \includegraphics[scale=0.58]{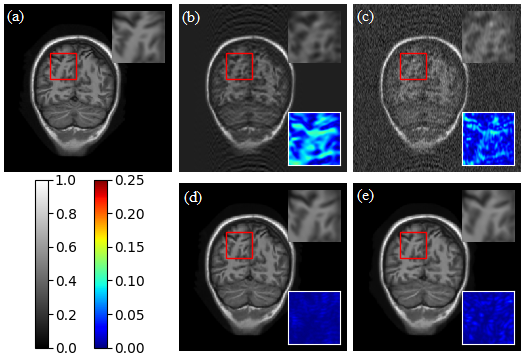}
\caption{Our method takes zero-filled reconstruction (ZFR) of the undersampled image as input and generates the corresponding reconstructed image. This can essentially be viewed as de-aliasing the ZFR. Example reconstruction results when 30\% data is retained. (a) Ground truth (GT), (b) ZFR of noise free image, (c) ZFR of image with 10\% noise, (d) results of the proposed method for noise-free image, and (e) results of the proposed method for image with 10\% noise. The top right inset indicates the zoomed in region of interest (ROI) corresponding to the red box, and the bottom right inset indicates the absolute difference between the ROI and the corresponding GT. The images are normalized between 0 and 1.
}
\label{intro_img}
\end{figure}

Compressive sensing (CS) \cite{donoho} can be used to accelerate the MRI acquisition process by undersampling the k-space data. Reconstruction of CS-MRI is an ill-posed inverse problem \cite{illposed}. Conventional CS-MRI frameworks assume prior information on the structure of MRI by making use of predefined sparsifying transforms such as the discrete wavelet transform, discrete cosine transform, etc. to obtain the solution \cite{lustig_sparse}. Instead of using predefined transforms, the sparse representation can be learnt from the data itself, i.e. dictionary learning (DLMRI) \cite{dlmri}. In \cite{bm3d}, a different approach of alternating between solving the optimization problem for reconstruction and denoising the image using block matching 3D (BM3D) model is adopted.  These frameworks however, suffer from the long computation time taken by iterative optimization processes \cite{iterative} as well as the assumption of sparse signals \cite{lustig_sparse}, which might not be able to fully capture the fine details \cite{sparsenotvalid}. 

Bora \textit{et al.} \cite{bora} have shown that instead of using the sparsity model, the CS signal can be recovered using pretrained generative models, where they use an iterative optimization to obtain the reconstructed signal. Another deep learning based approach was introduced by Yang \textit{et al.} \cite{deepadmm}, where a data flow graph is designed for alternating direction method of multipliers \cite{admm} to train the network (DeepADMM) for CS-MRI reconstruction. The inference phase takes a time similar to ADMM although the optimized parameters used are learned during the training process. A network architecture resembling a cascade of convolutional neural networks (CNNs) is proposed in \cite{cascade} (DeepCascade) which aims to reconstruct  dynamic sequences as well as independent frames of 2D MR images undersampled using Cartesian masks. The cascading network laid out resembles dictionary learning   reconstruction approaches, where the proposed approach can be viewed as an extended version of DLMRI. In \cite{deep_residual}, the authors unroll a residual learning approach where they use a deep CNN to learn the aliasing artifacts in the undersampled image, and subtract the aliasing artifacts thus estimated to obtain the de-aliased output.

Recent works \cite{dagan, stanford} demonstrate the application of generative adversarial networks (GANs) \cite{gan} to reconstruct CS-MRI. In these works, the use of a large set of CS-MR images and their fully sampled counterparts for training the GAN model can facilitate the extraction of prior information required to solve the reconstruction problem \cite{romberg}. The trained model is then used to obtain the reconstructed output for a new CS-MR image in a very short time. In \cite{dagan}, the authors propose a refinement learning based approach to obtain the de-aliased reconstructed MR image using a conditional GAN framework (DAGAN). Mardani \textit{et al.} \cite{stanford} (GANCS) use pixel-wise $\ell_1/\ell_2$ loss to train the generator and a least-squares GAN framework.

Many of the aforementioned works, including DeepCascade and DAGAN use $\ell_2$ loss function in the pixel domain for training, which is known to give blurry and excessively smooth outputs. Minimizing the $\ell_2$ or $\ell_1/\ell_2$ norm of the pixel-wise difference does result in a higher peak-signal-to-noise ratio (PSNR) of the reconstructed image, but it does not ensure good reconstruction of the structural details \cite{mse_leave}. In terms of frequency, the use of pixel-wise difference based loss mainly focuses on preserving low frequency components and does not enforce good reconstruction of high frequency details. Moreover, these state-of-the-art methods use discriminators which consider the input in a global sense while classifying. This may not allow the discriminator to consider the fine high frequency textural details, which are of vital importance in the MR images. Although DAGAN uses a frequency domain $\ell_2$ loss, it has the drawback of penalizing larger differences more, and allowing several smaller differences. This can yield a reconstructed output that looks similar to the ground truth but fails to preserve the finer details in the form of high frequency and structural content. 

\textit{\textbf{Contributions}}: To overcome these drawbacks, we incorporate the $\ell_1$ norm of the pixel-wise difference in the generator loss function to avoid blurry reconstruction. In order to preserve the structural and textural details in the reconstructed image, we propose the use of a structural similarity (SSIM) index based loss to train the generator. Moreover, to ensure better reconstruction of high frequency content in the MR images, we propose the use of a patch-based discriminator. Further, we propose a novel generator architecture by incorporating residual in residual dense blocks (RRDBs) in a U-net based architecture to utilize the benefits of residual and dense connections. It is also known that the binary cross-entropy based adversarial loss, which has been used in most of the previous works, makes the training of GANs unstable. Therefore, in order to stabilize the training process, we incorporate the Wasserstein loss. The reconstructed images should be less sensitive to the noise level in the measurements, since hardware devices are always susceptible to noise. In order to make the reconstruction robust to noise, we propose the use of noisy images for data augmentation to train our GAN model. Fig. \ref{intro_img} shows an example of reconstruction results obtained by the proposed approach (described in section 2 and 3) on noise-free as well as images contaminated with noise.







\section{Methodology}
\label{sec:pagestyle}
The acquisition model for the CS-MRI reconstruction problem in discrete domain can be described as:
\begin{equation}
    u = \mathbf{G}y + \mathbf{\eta},
\end{equation}
where $y\in\mathds{C}^{N^2}$ is a vector formed by the pixel values in the $N\times N$ desired image, $u\in\mathds{C}^{M}$ denotes the observation vector, and $\eta\in\mathds{C}^{M}$ is the noise vector. $\mathds{C}$ denotes the set of complex numbers. The matrix $\mathbf{G}$ describes the process of random undersampling in the k-space. It is the product of an ${N^2}\times{N^2}$ matrix $\mathbf{F}$, which computes the Fourier transform, and an ${M}\times{N^2}$ undersampling matrix $\textbf{U}$. Given an observation vector $u$, the reconstruction problem is to find out the corresponding $y$, considering $\eta$ to be a non-zero vector. We choose to find the solution to this reconstruction problem using a GAN model.

A GAN model comprises of a generator $G$ and a discriminator $D$, where the generator tries to fool the discriminator by transforming input vector $z$ to the distribution of true data $y_{true}$. On the other hand, the discriminator attempts to distinguish samples of $y_{true}$ from generated samples $G(z)$. 
\begin{figure*}
  \centering
\includegraphics[scale = 0.13]{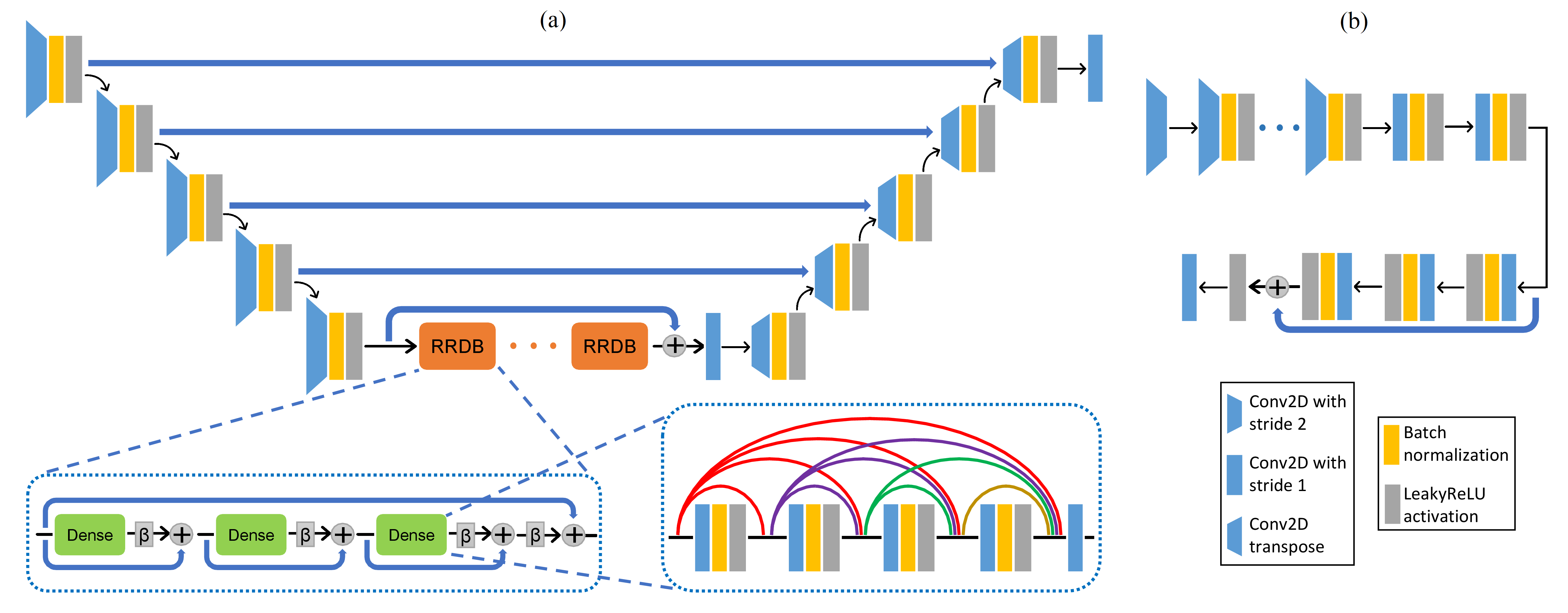}
\caption{(a) Generator architecture and  (b) discriminator architecture.}
\label{gen_disc}
\end{figure*}
We incorporate the conditional GAN (cGAN) based framework \cite{cgan} in our study. The model is conditioned on the aliased zero-filled reconstruction (ZFR) $x\in\mathds{C}^{N^2}$, given by $x = \mathbf{G}^Hu$, where $H$ denotes the Hermitian operator.  Instead of using a binary cross-entropy based adversarial loss for training the cGAN model, we use the Wasserstein loss \cite{wgan}. This helps in stabilizing the training process of standard GANs, which suffer from saturation resulting in vanishing gradients. Mathematically, the cGAN model with the Wasserstein loss solves the following optimization problem:
\begin{equation}
\begin{split}
    \min_{G}\max_{D}V_{WGAN} &= \E_{\mathbf{y}\sim p_y(\mathbf{y})}(D(y))\\
    &-\E_{\mathbf{x}\sim p_x(\mathbf{x})}(D(G(x))),
\end{split}
\end{equation}
where $V_{WGAN}$ denotes the value function and $\E$ denotes the expectation over a batch of images. $p_y(\mathbf{y})$ and $p_x(\mathbf{x})$ denote the distribution of GT and ZFR images, respectively. The optimization problem is solved by alternating between $p$ steps where discriminator ($D$) is optimized and a single step of generator ($G$) optimization. The loss function which is minimized while training the discriminator is given by:
\begin{equation}
    L_{DIS} = \E_{\mathbf{x}\sim p_x(\mathbf{x})}(D(G(x))) - \E_{\mathbf{y}\sim p_y(\mathbf{y})}(D(y)).
\end{equation}
The Lipschitz constraint is enforced by applying weight clipping on the weights of the discriminator \cite{wgan}.

Fig. \ref{gen_disc} (a) shows the generator architecture of the proposed model. The architecture is based on a U-net \cite{unet}, which consists of several encoders and corresponding decoders. Each encoder is in the form of a convolutional layer, which decreases the size and increases the number of feature maps. Each decoder consists of a transposed convolutional layer, to increase the size of the feature maps. In order to transfer the features of a particular size from the encoder to the corresponding decoder, skip connections are present. Instead of obtaining feature maps of size lower than $\frac{N}{32}\times\frac{N}{32}$ using more encoders (and decoders), the proposed architecture consists of RRDBs at the bottom of the U-net. The addition of RRDBs at the bottleneck layer helps in increasing the depth of the network which can enable learning of more complicated functions. Each RRDB \cite{rrdb} consists of dense blocks, as well as residual connections at two levels: across each dense block, and across all the dense blocks in one RRDB, as shown in Fig. \ref{gen_disc} (a). The output of each dense block is scaled by $\beta$ before it is added to the identity mapping. Residual connections make the length of the network variable thereby making identity mappings easier to learn and avoid vanishing gradients in the shallower layers. Dense connections allow the transfer of feature maps to deeper layers, thus increasing the variety of accessible information. Just like residual connections, they also help in alleviating vanishing gradients. Moreover, their use reduces the number of parameters as compared to conventional convolutional networks, since the necessity to learn redundant information has been removed.  Throughout this network, batch normalization (BN) and leaky rectified linear unit (ReLU) activation are applied after each convolutional layer. At the output, a hyperbolic tangent activation is used. 

The discriminator is a CNN with 11 layers, as illustrated in Fig. \ref{gen_disc} (b). Each layer consists of a convolutional layer, followed by BN and leaky ReLU activation.  A patch-based discriminator \cite{patchgan} is
 incorporated in order to improve the preservation of high frequency details in the reconstructed output, since $\ell_1$ norm of the pixel-wise difference (used as a loss function in this work) mainly focuses on preservation of low frequency components and does not enforce good reconstruction of high frequency details. This discriminator focuses on the local patches, tries to score each patch (size $m \times m$) of the image separately in an attempt to classify whether the patch is real or fake, and gives the average score as the final output.

In order to reduce the pixel-wise difference between the generated image and the corresponding ground truth (GT) image, a mean absolute error (MAE) based loss is incorporated while training the generator. It is given by:
\begin{equation}
    L_{MAE} = \E(\|G(x)-y\|_1),
\end{equation}
where $\|\cdot\|_1$ denotes the $\ell_1$ norm. Since the human vision system is sensitive to structural distortions in images, it is important to preserve the structural information in MR images, which is crucial for clinical analysis. Moreover, $\ell_1$ norm minimization of the pixel-wise difference does not enforce textural and structural correctness,  which may lead to a reconstructed output of poor diagnostic quality. Super resolution is another well-known inverse problem that tries to interpolate both low frequency and high frequency components from a low resolution image. Inspired by previous works on super resolution \cite{nvidia}, a mean SSIM (mSSIM) \cite{ssim} based loss is incorporated in order to improve the reconstruction of fine textural details in the images. It is formulated as follows:
\begin{equation}
    L_{mSSIM} = 1-\E\left(\frac{1}{K}\sum_{j=1}^{K}SSIM(G_j(x),y_j)\right),
\end{equation} 
where $K$ is the number of patches in the image, and SSIM is calculated as follows: 
\begin{equation}
    SSIM(u,v)=\frac{2\mu_u\mu_v+c_1}{\mu_u^2+\mu_v^2+c_1}\frac{2\sigma_{uv}+c_2}{\sigma_u^2+\sigma_v^2+c_2},
\end{equation}
where $u$ and $v$ represent two patches, and $\mu$ and $\sigma$ denote the mean and variance, respectively. $c_1$ and $c_2$ are small constants to avoid division by zero. 

The overall loss for training the generator is given by:
\begin{equation}
    L_{GEN} = \alpha_1 L_{MAE} + \alpha_2 L_{mSSIM} - \alpha_3 \E(D(G(x))),
\end{equation}
where $\alpha_1$, $\alpha_2$, and $\alpha_3$ are the weighting factors for various loss terms.


\section{Results and Discussion}
\subsection{Training settings} In this work, a 1-D Gaussian mask is used for undersampling the k-space. Since the ZFR $x$ is complex valued, the real and imaginary components are concatenated and passed to the generator in the form of a two channel real valued input. The batch size is set as 32. The discriminator is updated three times before every generator update. The threshold for weight clipping is 0.05. The growth rate for the dense blocks is set as 32, $\beta$ is 0.2, and 12 RRDBs are used. The number of filters in the last layer of each RRDB is 512. Adam optimizer \cite{adam} is used for training with $\beta_1 = 0.5$ and $\beta_2=0.999$. The learning rate is set as $10^{-4}$ for the generator and $2\times10^{-4}$ for the discriminator. The weighting factors are $\alpha_1 = 20$, $\alpha_2 = 1$, and $\alpha_3 = 0.01$. The model \footnote{The code is available at: \url{https://puneesh00.github.io/cs-mri/}} is implemented using Keras framework \cite{keras} with TensorFlow backend. For training, 4 NVIDIA GeForce GTX 1080 Ti GPUs are used, each having 11 GB RAM.

\begin{table*}[h!]
\centering
\caption{Ablation study of the model.}
\label{ablation}
\resizebox{1.6\columnwidth}{!}{
\begin{tabular}{|c|c|c|c|c|c|} 
 \hline
 \rule{0pt}{9pt} Network Settings & $1^{st}$ & $2^{nd}$ & $3^{rd}$ & $4^{th}$ & $5^{th}$\\
 \hline
\rule{0pt}{9pt} U-net G + patch-based D & \cmark & \cmark & \cmark & \cmark & \cmark \\ 
\rule{0pt}{9pt} RRDBs & \xmark & \cmark & \cmark & \cmark & \cmark \\
\rule{0pt}{9pt} BN in RRDBs & \xmark & \xmark & \cmark & \cmark & \cmark \\
\rule{0pt}{9pt} Data augmentation & \xmark & \xmark & \xmark & \cmark & \cmark \\
\rule{0pt}{9pt} Wasserstein loss & \cmark & \cmark & \cmark & \xmark & \cmark \\
 \hline
 \hline
  \rule{0pt}{9pt} Images & \multicolumn{5}{c|}{PSNR (dB) / mSSIM} \\
  \hline
 \rule{0pt}{9pt} Noise-free & 40.45 / 0.9865 & 41.39 / 0.9810 & 41.88 / 0.9829 & 41.80 / 0.9820 & 42.31 / 0.9841\\

 \rule{0pt}{9pt} 10\% noise added & 38.25 / 0.9641 & 38.03 / 0.9624 & 38.03 / 0.9620 & 39.55 / 0.9728 & 39.80 / 0.9751\\
 
 \rule{0pt}{9pt} 20\% noise added & 33.98 / 0.9217 & 34.01 / 0.9210 & 33.78 / 0.9180 & 37.21 / 0.9576 & 37.56 / 0.9619\\

 \hline
\end{tabular}}
\end{table*}
\begin{figure*}
\center
    \includegraphics[scale=0.7]{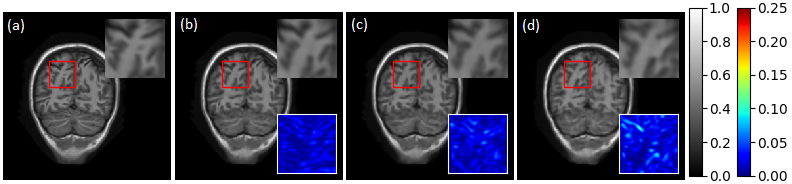}
\caption{Reconstruction results of the proposed method for 20\% undersampled images, taken from the MICCAI dataset. (a) GT, reconstruction results for (b) noise-free image, (c) image with 10\% noise, and (d) image with 20\% noise. The top right inset indicates the zoomed in ROI corresponding to the red box. The bottom right inset indicates the absolute difference between the ROI and the corresponding GT. The images are normalized between 0 and 1.
}
\label{res_a5}
\end{figure*}

\subsection{Data details} For the purpose of training and testing, two different datasets are used. We first evaluate our model on T-1 weighted MR images of brain from the MICCAI 2013 grand challenge dataset \cite{dataset}. This is followed by another evaluation using MR images of knee (coronal view) from the MRNet dataset \cite{mrnet}. The images in both the datasets are of size $256\times256$. In order to make the reconstructed output robust to noise, data augmentation is carried out using images with 10\% and 20\% additive complex Gaussian noise in the k-space. To make the set of training images for the MICCAI dataset, 19\,797 images are randomly taken from the training set of the aforementioned dataset. Out of these, noise is added to 6335 images, while the remaining 13\,462 images are used without any noise. In addition, 990 images are chosen from the 13\,462 noise-free images, and noise is added to them also, to get a total of 20\,787 images for training. Among the noisy images, number of images with 10\% and 20\% noise is equal. Thus, the set contains 64.76\% noise-free images, 30.48\% noisy images whose corresponding noise-free images are not present in the training set, and 4.76\% noisy images whose corresponding noise-free images are present in the training set. To make the set of training images for the MRNet dataset, a total of 12\,500 images are taken from the training set, where the aforementioned ratio of noise-free, overlapping noisy images, and non-overlapping noisy images is maintained. For testing, 2000 images are chosen randomly from the test sets of the respective datasets. The tests are conducted in three stages: using noise-free images, using images with 10\% noise added to them, and using images with 20\% noise.

\subsection{Results}
Table \ref{ablation} summarizes the quantitative results to study the effect of addition of various components to the model. These results are reported for images taken from the MICCAI dataset, in which 20\% of the raw k-space samples are retained. For all the cases, the generator is trained with $L_{GEN}$. In the first case, the GAN model comprises of a U-net generator (without RRDBs) and a patch-based discriminator, with BN present throughout the network. It is trained with Wasserstein loss. In the subsequent cases, the use of RRDBs (without BN), followed by addition of BN to RRDBs results in significant improvement in PSNR of the reconstructed outputs corresponding to noise-free images. As mentioned in Section 3.1, the loss function used in training of all the networks takes the weight for $L_{MAE}$ as $20$ times the weight for $L_{mSSIM}$. Such a ratio might cause the model to focus more towards reducing the MAE. This results in a more consistent performance of PSNR as the training progresses. In the inference on noisy test images (both 10\% and 20\%), the PSNR and mSSIM have relatively less consistent performance as seen in the ablation studies. One possible reason for this observation might be the large number of nonlinearities present in the model, which give the ability to learn a highly complex function. As mentioned in \cite{degrads}, a highly complex function can have improved performance for the noise-free case at the cost of slightly increased sensitivity to noise as compared to its less complex counterparts. The use of data augmentation with noisy images, in the fifth case, results in significantly better quantitative results for the reconstruction of noisy images, as compared to the first three cases. This improves the robustness of the model. In the fourth case, we train the network with the conventional binary cross-entropy based adversarial loss instead of Wasserstein loss. On comparing this case with the fifth case, it is evident that the use of Wasserstein loss improves the training process. The settings of the fifth case are finalized and used for the subsequent results reported in this work.

\begin{figure*}
\center
    \includegraphics[scale=0.62]{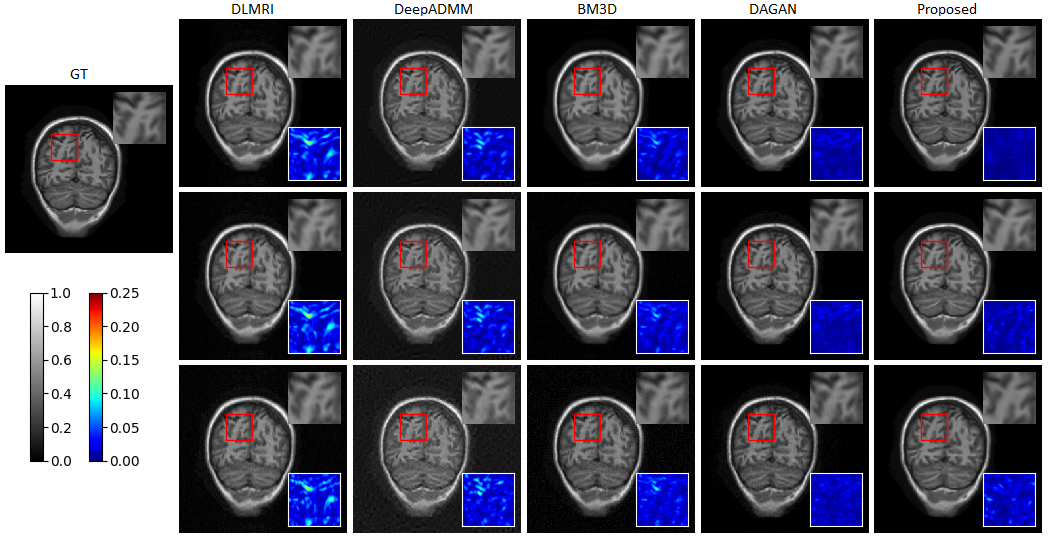} \caption{Qualitative results and comparison with previous methods for 30\% undersampled images, taken from the MICCAI dataset. The first row shows reconstruction results for noise-free images, the second row shows reconstruction results for images with 10\% noise, and the third row shows reconstruction results for images with 20\% noise. The top right inset indicates the zoomed in ROI corresponding to the red box. The bottom right inset indicates the absolute difference between the ROI and the corresponding GT. The images are normalized between 0 and 1.}
\label{comp_a3}
\end{figure*}

\begin{table*}[h!]
\centering
\caption{Quantitative comparison with previous methods using MICCAI dataset.}
\label{tab_comp}
\resizebox{1.7\columnwidth}{!}{
\begin{tabular}{|c|c|c|c|c|c|c|c|} 
 \hline
 \rule{0pt}{9pt}\multirow{2}{*}{Method} & \multicolumn{2}{c|}{Noise-free images} & \multicolumn{2}{c|}{10\% noise added} & \multicolumn{2}{c|}{20\% noise added} & Reconstruction/ \\ 
\cline{2-7}
 \rule{0pt}{9pt}  & PSNR (dB) & mSSIM & PSNR (dB) & mSSIM & PSNR (dB) & mSSIM &  Test time (s)\\
 \hline
 \rule{0pt}{9pt} DLMRI\cite{dlmri} & 37.405 & 0.8732 & 34.144 & 0.6140 & 31.564 & 0.4346 & 25.7732\\
 \rule{0pt}{9pt} DeepADMM\cite{deepadmm} & 41.545 & 0.8946 & 39.078 & 0.8105 & 35.373 & 0.6000 & 0.3135\\
  \rule{0pt}{9pt} BM3D\cite{bm3d} & 42.521 & 0.9764 & 37.836 & 0.7317 & 33.657 & 0.4947 & 6.8230\\
 \rule{0pt}{9pt} DAGAN\cite{dagan} & 43.329 & 0.9860 & 42.006 & 0.9814 & 39.160 & 0.9619 & \textbf{0.0063}\\
 \rule{0pt}{9pt} Proposed & \textbf{46.877} & \textbf{0.9943} & \textbf{42.338} & \textbf{0.9855} & \textbf{39.493} & \textbf{0.9740} & 0.0091\\
 \hline
\end{tabular}}
\end{table*}

The qualitative results of the proposed method are shown in Fig. \ref{res_a5} for 20\% undersampled images taken from the MICCAI dataset. It can be seen that the proposed method is able to reconstruct the structural content in the image, including many fine details, successfully. This is also indicated by the quantitative results shown in Table \ref{ablation}. Also, the contrast of the reconstructed image looks very similar to that of the GT. The reconstruction results for noisy inputs, as well as their differences with the corresponding GT, indicate the robustness of the model.

Fig. \ref{comp_a3} and Table \ref{tab_comp} show the qualitative and quantitative comparison of the proposed method, respectively, with some state-of-the-art methods like DLMRI \cite{dlmri}, DeepADMM \cite{deepadmm}, BM3D \cite{bm3d}, and DAGAN \cite{dagan}. These results are reported for images taken from the MICCAI dataset, in which 30\% of the k-space data is retained. The comparison of the zoomed in ROI of the reconstructed outputs corresponding to the noise-free images, produced by the aforementioned methods, as well as the difference with the GT show that these methods are not able to fully preserve the structural content present in the GT. It can be seen that our method produces the least difference between the ROI and the corresponding GT. Even in the case of noisy images, our method is robust to the artifacts in the image as it produces a smooth background, similar to the GT, whereas other methods produce outputs with noisy artifacts as well as granularity. This can be seen in the results shown in the second and third row in Fig. \ref{comp_a3}. The artifacts are more visible in the background of the zoomed in ROI, whereas the granularity can be more easily seen in the greyish boundary surrounding the brain structure as well as in the black background. Moreover, the contrast is much better preserved in our reconstructed outputs as seen in the zoomed in ROI of all the three rows in Fig. \ref{comp_a3}. 

The quantitative results also reinforce the effectiveness of the proposed method. Table \ref{tab_comp} shows that both the PSNR and mSSIM for the proposed method are significantly better than the previous methods for noise-free as well as images with 10\% and 20\% noise. All the previous methods, with the exception of DAGAN, experience a significant decline in the PSNR and mSSIM values when their reconstruction results for noise-free and noisy images are compared. This proves that the reconstruction quality significantly deteriorates on addition of noise as these methods lack robustness. It is observed that the proposed method significantly outperforms the other methods in the noise-free setting, but the improvement in the noisy setting is less significant. As mentioned before, this might be the result of the large number of nonlinearities present in the model, which allow the learned function to be highly complex and obtain better performance for the noise-free case at the cost of slightly more sensitivity to noise \cite{degrads}. However, the proposed augmentation technique increases the robustness of the model, as seen by the results presented in Table \ref{tab_comp}. Moreover, the reconstruction time of the proposed method is 9.06 ms per image, which can facilitate real-time reconstruction of MR images. DLMRI and BM3D have a much higher reconstruction time owing to the iterative fashion in which they obtain the output. On the other hand, GAN based approaches have a reconstruction time of the order of milliseconds as the testing phase only involves a forward pass through the trained generator. For the generator model of DAGAN and the proposed approach, the FLOPs (total number of  floating-point operations) are 197M and 314M, respectively. The corresponding FLOPS (floating-point operations per second) are 31.31G and 34.55G, which are calculated using the inference times listed in Table \ref{tab_comp}. 

\begin{figure}[h!]
\center
    \includegraphics[scale=0.59]{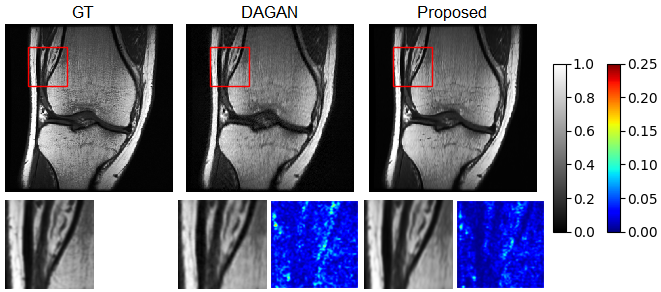}
\caption{Qualitative results and comparison for 30\% undersampled image, taken from the MRNet dataset. These are the reconstruction results for noise-free images. The left inset indicates the zoomed in ROI corresponding to the red box. The right inset indicates the absolute difference between the ROI and the corresponding GT. The images are normalized between 0 and 1.
}
\label{knee_res}
\end{figure}

To demonstrate the generalization of the proposed approach, Table \ref{tab_comp2} and Fig. \ref{knee_res} show the quantitative and qualitative results for images taken from the MRNet dataset, in which 30\% of the k-space data is retained, as well as the comparison with DAGAN \cite{dagan}, which obtained the closest results on the MICCAI dataset. From Table \ref{tab_comp2}, it is observed that the proposed method outperforms DAGAN by a significant margin as it obtains better results for images with 20\% noise than those obtained by DAGAN on noise-free images. It is also observed that the PSNR and mSSIM values obtained for MRNet dataset are lower than those obtained for MICCAI dataset. One possible reason for this might be the larger black region present in the images in the MICCAI dataset,  which lacks any details or structural information. Fig. \ref{knee_res} shows that the proposed method is able to obtain a reconstructed output of high quality, as the difference between the GT and the reconstructed image is very low. 

\begin{table}[h!]
\centering
\caption{Quantitative results and comparison using MRNet dataset.}
\label{tab_comp2}
\resizebox{0.99\columnwidth}{!}{
\begin{tabular}{|c|c|c|c|c|c|c|} 
 \hline
 \rule{0pt}{9pt}\multirow{2}{*}{Method}  & \multicolumn{6}{c|}{PSNR (dB) / mSSIM} \\ 
\cline{2-7}
 \rule{0pt}{9pt} & \multicolumn{2}{c|}{Noise-free images} & \multicolumn{2}{c|}{10\% noise added} & \multicolumn{2}{c|}{20\% noise added}  \\
 \hline

 \rule{0pt}{9pt} DAGAN\cite{dagan} & \multicolumn{2}{c|}{31.529 / 0.8754} & \multicolumn{2}{c|}{30.452 / 0.8182} & \multicolumn{2}{c|}{28.267 / 0.7098} \\
 \rule{0pt}{9pt} Proposed & \multicolumn{2}{c|}{\textbf{34.823} / \textbf{0.9412}} & \multicolumn{2}{c|}{\textbf{33.522} / \textbf{0.9167}} & \multicolumn{2}{c|}{\textbf{32.034} / \textbf{0.8884}} \\
 \hline
\end{tabular}}
\end{table}

\begin{figure}[h!]
\center
    \includegraphics[scale=0.55]{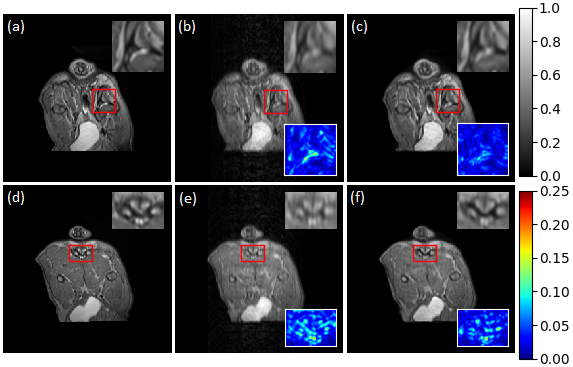}
\caption{Results of zero-shot inference. (a,d) GT, (b,e) ZFR, (c,f) reconstruction results for noise-free image. The top right inset indicates the zoomed in ROI corresponding to the red box. The bottom right inset indicates the absolute difference between the ROI and the corresponding GT. The images are normalized between 0 and 1.
}
\label{zerosh}
\end{figure}

We also tested the model trained on MR images of brain from the MICCAI dataset to reconstruct MR images of canine legs from the MICCAI 2013 challenge. Fig. \ref{zerosh} shows the results of this zero-shot inference for images in which 20\% of the k-space data is retained. Though no images of canine legs were used for training, the model is able to faithfully reconstruct most of the structural content, and is able to achieve average PSNR and mSSIM values of 41.28 dB and 0.9788, respectively, for 2000 test images.

Further, we performed the zero-shot inference of the model trained on 30\% undersampled MR images of knee from the MRNet dataset to reconstruct MR images of canine legs from the MICCAI 2013 challenge. It is able to achieve average PSNR and mSSIM values of 43.79 dB and 0.9883, respectively, for 2000 test images.

\textbf{\textit{Potential hallucination by GANs}}: Conventional GAN training techniques may suffer from hallucination of details which could potentially be harmful for image diagnosis. The proposed scheme tries to control the hallucination of details by the use of pixel-wise MAE loss as well as the mSSIM based loss in the image domain, both of which try to ensure that the generated image is close to the GT. $L_{MAE}$ tries to make sure that the low frequency details are of the generated output as closely aligned to the ground truth, whereas $L_{mSSIM}$ focuses on increasing the similarity of generated and GT in terms of structural details. The zero-shot inference is also helpful in pointing out that the proposed GAN model has shown no sign of hallucination on data samples taken from a distribution that is different from the training distribution.

\begin{figure}[h!]
\center
    \includegraphics[scale=0.52]{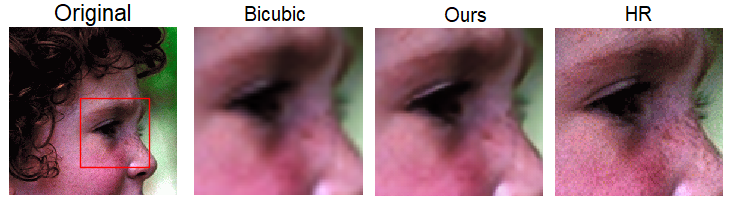}
\caption{Qualitative results of the SR experiment on an image from Set5.
}
\label{head}
\end{figure}

\begin{figure}[h!]
\center
    \includegraphics[scale=0.41]{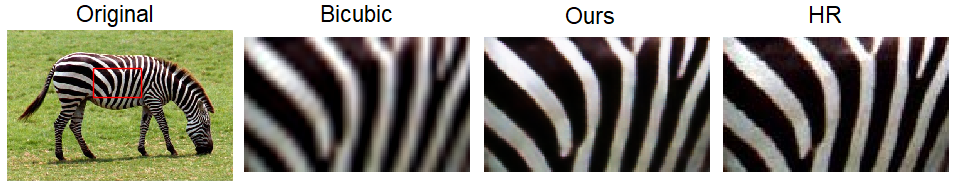}
\caption{Qualitative results of the SR experiment on an image from Set14.
}
\label{zebra}
\end{figure}

\textbf{\textit{Additional Experiment for Super Resolution}}: To understand the usability of the proposed model in other computer vision applications, we take super resolution (SR) as a vision task, for which the model is not optimized, and evaluate its performance on some commonly used datasets for this task. All the hyperparameter values mentioned in section 3.1 are maintained for this experiment, except $\alpha_1$, which is set as 30. For training, patches of size $192\times192$ were used from the images present in the DIV2K \cite{div2k} and the Flickr2K datasets. These are the high resolution (HR) or the GT images ($y$). This variable image size is supported by the fully convolutional architecture of the generator as well as the discriminator, i.e. neither of the networks involve the use of dense layers. The patch-based discriminator, which classifies sections of the image and not the entire image, also allows the image size to be variable. For the task of $4\times$ super resolution, the corresponding low resolution (LR) images of size $48\times48$ are used. These are obtained using bicubic downsampling, which is a widely used degradation model. As the size of the input and the output images is the same in our framework, we use the images obtained using bicubic interpolation on the low resolution images as the input for the model ($x$). For the purpose of testing, we use Set5 \cite{set5} and Set14 \cite{set14} datasets. The quantitative results of our model as well as the comparison with previous methods for $4\times$ super resolution are presented in Table \ref{tab_SR}. The PSNR and mSSIM values are calculated by considering only the Y channel of the images, after converting them from RGB to YCbCr colorspace, as mentioned in several previous works on super resolution. The qualitative results obtained using our approach are illustrated in Fig. \ref{head} and Fig. \ref{zebra}.  Although the proposed framework is optimized for the task of CS-MRI reconstruction, it gives satisfactory performance for the super resolution task as well.

\begin{table}[h!]
\centering
\caption{Quantitative results and comparison with previous methods for the SR experiment.}
\label{tab_SR}
\resizebox{0.75\columnwidth}{!}{
\begin{tabular}{|c|c|c|c|c|} 
 \hline
 \rule{0pt}{9pt}\multirow{2}{*}{Method}  & \multicolumn{4}{c|}{PSNR (dB) / mSSIM} \\ 
\cline{2-5}
 \rule{0pt}{9pt} & \multicolumn{2}{c|}{Set5} & \multicolumn{2}{c|}{Set14} 
 \\
 \hline

 \rule{0pt}{9pt} Bicubic & \multicolumn{2}{c|}{28.42 / 0.8104} & \multicolumn{2}{c|}{26.00 / 0.7027} 
 \\
 \rule{0pt}{9pt} SRCNN\cite{srcnn} & \multicolumn{2}{c|}{30.48 / 0.8628} & \multicolumn{2}{c|}{27.50 / 0.7513} 
 \\
 
 \rule{0pt}{9pt} VDSR\cite{vdsr} & \multicolumn{2}{c|}{31.35 / 0.8830} & \multicolumn{2}{c|}{28.02 / 0.7680}  
 \\
 
 \rule{0pt}{9pt} FSRCNN\cite{fsrcnn} & \multicolumn{2}{c|}{30.72 / 0.8660} & \multicolumn{2}{c|}{27.61 / 0.7550}  
 \\
 
 \rule{0pt}{9pt} LapSRN\cite{LapSRN} & \multicolumn{2}{c|}{31.54 / 0.8850} & \multicolumn{2}{c|}{28.19 / 0.7720}  
 \\
  \rule{0pt}{9pt} EDSR\cite{edsr} & \multicolumn{2}{c|}{32.46 / 0.8968} & \multicolumn{2}{c|}{28.80 / 0.7876}  
 \\
 \rule{0pt}{9pt} Ours & \multicolumn{2}{c|}{31.63 / 0.8982} & \multicolumn{2}{c|}{27.62 / 0.7784}  
 \\
 \hline
\end{tabular}}
\end{table}

\section{Conclusion}
\label{sec:prior}

In this paper, a novel GAN based framework has been utilized for CS-MRI reconstruction. The use of RRDBs in a U-net based generator architecture increases the amount of information available. In order to preserve the high frequency content as well as the structural details in the reconstructed output, a patch-based discriminator and structural similarity based loss have been incorporated. The use of noisy images during training makes the reconstruction results highly robust to noise. The proposed method is able to outperform the state-of-the-art methods, while maintaining the feasibility of real-time reconstruction. In future, we plan to analyze the performance of the proposed model for different k-space sampling patterns. In order to improve the reconstruction time, we plan to work on lightweight architectures. Further work may be carried out on devising regularization terms that help to preserve the finest of details in the reconstructed output. 

\section*{Acknowledgement}
We would like to thank the reviewers for their valuable feedback and constructive comments.

{\small
\bibliographystyle{ieee_fullname}
\bibliography{egbib}
}

\end{document}